\documentclass[%
 reprint,
 amsmath,amssymb,
 aps,
]{revtex4-2}

\usepackage{graphicx}
\usepackage{dcolumn}
\usepackage{bm}

\usepackage{amsmath}
\usepackage{amssymb}
\usepackage{graphicx}
\usepackage{url}
\usepackage{epsfig}
\usepackage[T1]{fontenc}
\usepackage[latin9]{inputenc}
\usepackage{multirow}
\usepackage{xspace}
\usepackage{hyperref}
\usepackage{xcolor}
\usepackage{cprotect}
\usepackage{mathtools}
\usepackage{float}

\newcommand{\angantyr}{Angantyr\xspace}
\newcommand{\pythia}{P\protect\scalebox{0.8}{YTHIA}\xspace}
\newcommand{\pytppp}{P\protect\scalebox{0.8}{YTHIA}8\xspace}
\newcommand{\Pom}{\mathbb{P}}

\def\beq{\begin{equation}}
\def\eeq{\end{equation}}
\def\bear{\begin{eqnarray}}
\def\eear{\end{eqnarray}}
\def\pA{\ensuremath{{\rm p}A}}

\begin{document}

\preprint{LU-TP-19-11\\MCnet-19-29\\arXiv:1912.08733}

\title{Four-jet double parton scattering production in proton-nucleus collisions within the \pytppp framework}

\author{Oleh Fedkevych}
\email{oleh.fedkevych@ge.infn.it}
\affiliation{%
 Dipartimento di Fisica, Universit\`{a} di Genova, Via Dodecaneso 33, 16146 	Genoa, Italy
}%

\author{Leif L\"onnblad}
\email{leif.lonnblad@thep.lu.se} 
\affiliation{%
 Department of Astronomy and Theoretical Physics, S\"{o}lvegatan 14A, S-223 62 Lund, Sweden
}%

\date{\today}

\begin{abstract}
  We present our studies of four-jet double parton scattering
  production in proton-nucleus collisions within the framework of the
  \pytppp event generator. We demonstrate that double absorptive
  processes in p$A$ generated by the \angantyr model in \pytppp give an
  enhancement of the total double parton scattering cross section
  similar to the predictions by Strikman and Treleani in 2001.
  Additionally, we discuss how the growth of activity in a direction
  of a nucleus affects an $A$-scaling of a total double parton
  scattering cross section in proton-nucleus collisions.
\end{abstract}

\keywords{Double parton scattering, Hard scattering in relativistic heavy ion collisions, Monte Carlo simulations}
\maketitle


\section{\label{s:intro} Introduction}
Despite a significant progress in both theoretical and experimental
studies of QCD, many of its aspects still require further detailed
investigation.  One of the possible keys to a deeper understanding of
QCD and a structure of hadrons is the study of so called
\textit{double parton scattering} (DPS), a process when two hard
interactions occur in a single \textit{hadron-hadron}
collision. Various studies of DPS performed at
\mbox{\textit{proton-proton} (pp)} and
\mbox{\mbox{\textit{proton-antiproton}}} (p$\bar{\rm p}$) colliders
\cite{Akesson:1986iv, Alitti:1991rd, Abe:1993rv, Abe:1997bp,  Abe:1997xk, Abazov:2009gc, Aaboud:2016fzt, Aaij:2012dz, Aad:2013bjm, Chatrchyan:2013xxa, Abazov:2014qba, Abazov:2014fha, Aad:2014kba, Aaij:2015wpa, Abazov:2015fbl, Abazov:2015nnn, Aaboud:2016dea,  Aaij:2016bqq, Khachatryan:2016ydm, Aaboud:2018tiq, Sirunyan:2017hlu} suggest the presence of partonic correlations
which leads to small values of the effective DPS interaction area,
$\sigma_{eff}$. The nature of these correlations is still under debate
and is obscured by the difficulties involved with disentangling
different sources of parton correlations. As a tool for gaining
further insights, Strikman and Treleani proposed to study DPS
processes in \textit{proton-nucleus} (p$A$) collisions
\cite{Strikman:2001gz} which would allow the separation of
\textit{transverse} from \textit{longitudinal} parton correlations
according to the different $A$-dependence of the corresponding
contributions to a total DPS cross section. This idea got a further
development in \cite{Frankfurt:2004kn, Blok:2012jr, Strikman:2010bg, Blok:2017alw, Alvioli:2019kcy} and found some
phenomenological applications in a series of works
\cite{Frankfurt:2004kn,Blok:2012jr, dEnterria:2013mrp, Strikman:2010bg, Calucci:2013pza, Cattaruzza:2004qb, dEnterria:2012jam, Cazaroto:2013fua, dEnterria:2014lwk, Helenius:2019uge, Blok:2019fgg}.

While a significant progress in a theoretical description of DPS in
p$A$ collisions have been achieved, a framework for realistic
simulations of DPS in p$A$ collisions has been lacking. In this paper
we compare predictions of the Strikman $\&$ Treleani model against
predictions of the \angantyr model of p$A$ collisions
\cite{Bierlich:2018xfw} recently implemented in the \pytppp event
generator \cite{Sjostrand:2006za,Sjostrand:2007gs}.  We discuss in
detail the differences and similarities between the models and
demonstrate that the \angantyr model gives predictions similar to
those given by Strikman and
Treleani. Since \angantyr
takes advantage of the entire \pythia machinery, including
\textit{multiple parton interaction} (MPI), initial and final state
radiation and many other effects, we therefore conclude that it can be
used to give realistic Monte Carlo simulations of complete four-jet
DPS production events in p$A$ collisions.

This paper is organised as follows: in
Section~\ref{s:strikman_treleani_model} and
Section~\ref{s:angantyr_model} we briefly sketch out the \mbox{Strikman $\&$
  Treleani} and \angantyr models respectively, in
Section~\ref{s:predictions_of_pythia} we provide our simulations for
the four-jet DPS production in p$A$ collisions and compare them against
predictions made within the framework of Strikman and Treleani and in
Section~\ref{s:conclusions} we summarise our results and discuss some
further perspectives of DPS modelling in p$A$ collisions.

\section{\label{s:strikman_treleani_model} Strikman $\&$ Treleani model}
\begin{figure}
\centering
\begin{minipage}[h]{0.44\linewidth}
\center{\includegraphics[width=0.9\linewidth]{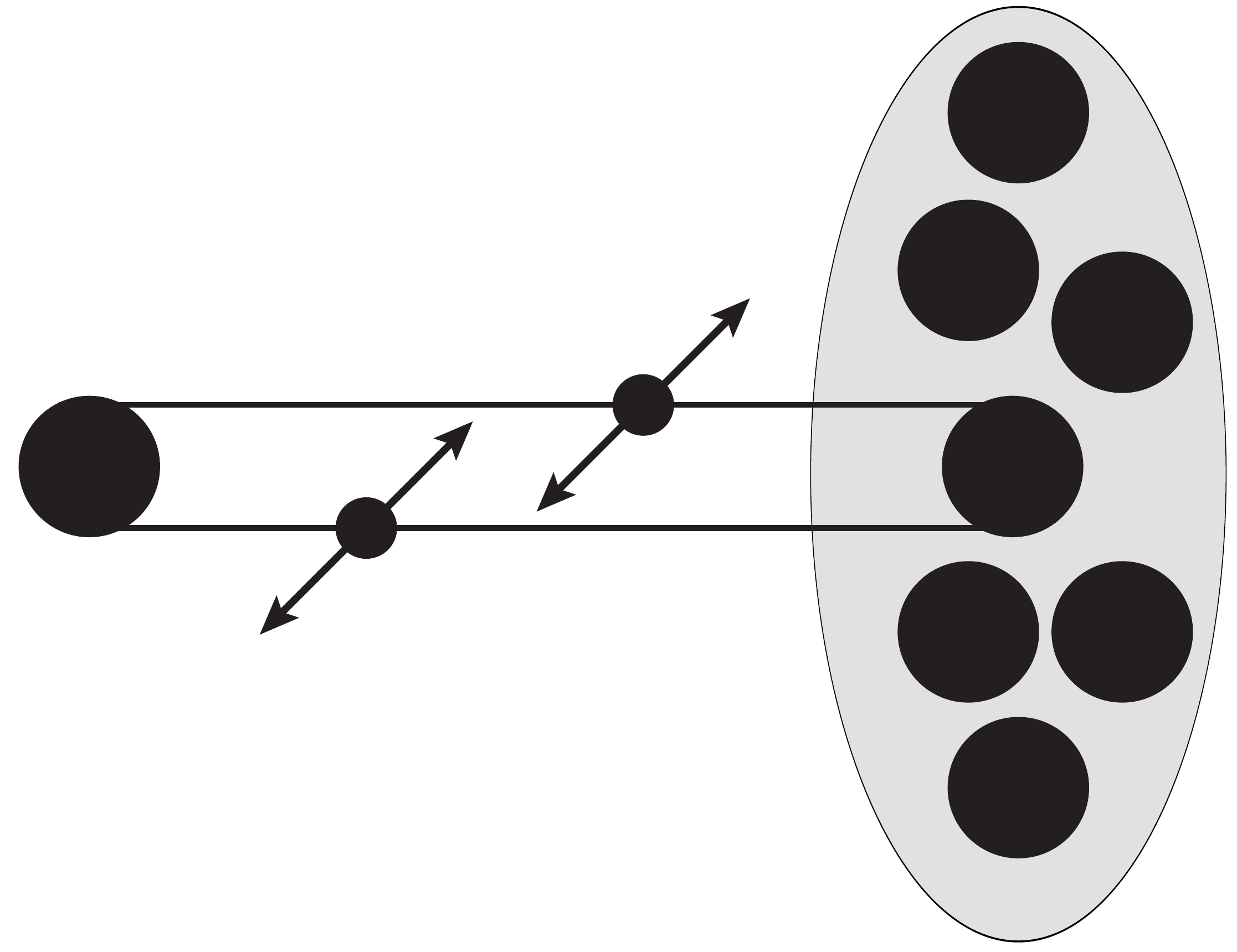}} \\a)
\end{minipage}
\hfill
\begin{minipage}[h]{0.44\linewidth}
\center{\includegraphics[width=0.9\linewidth]{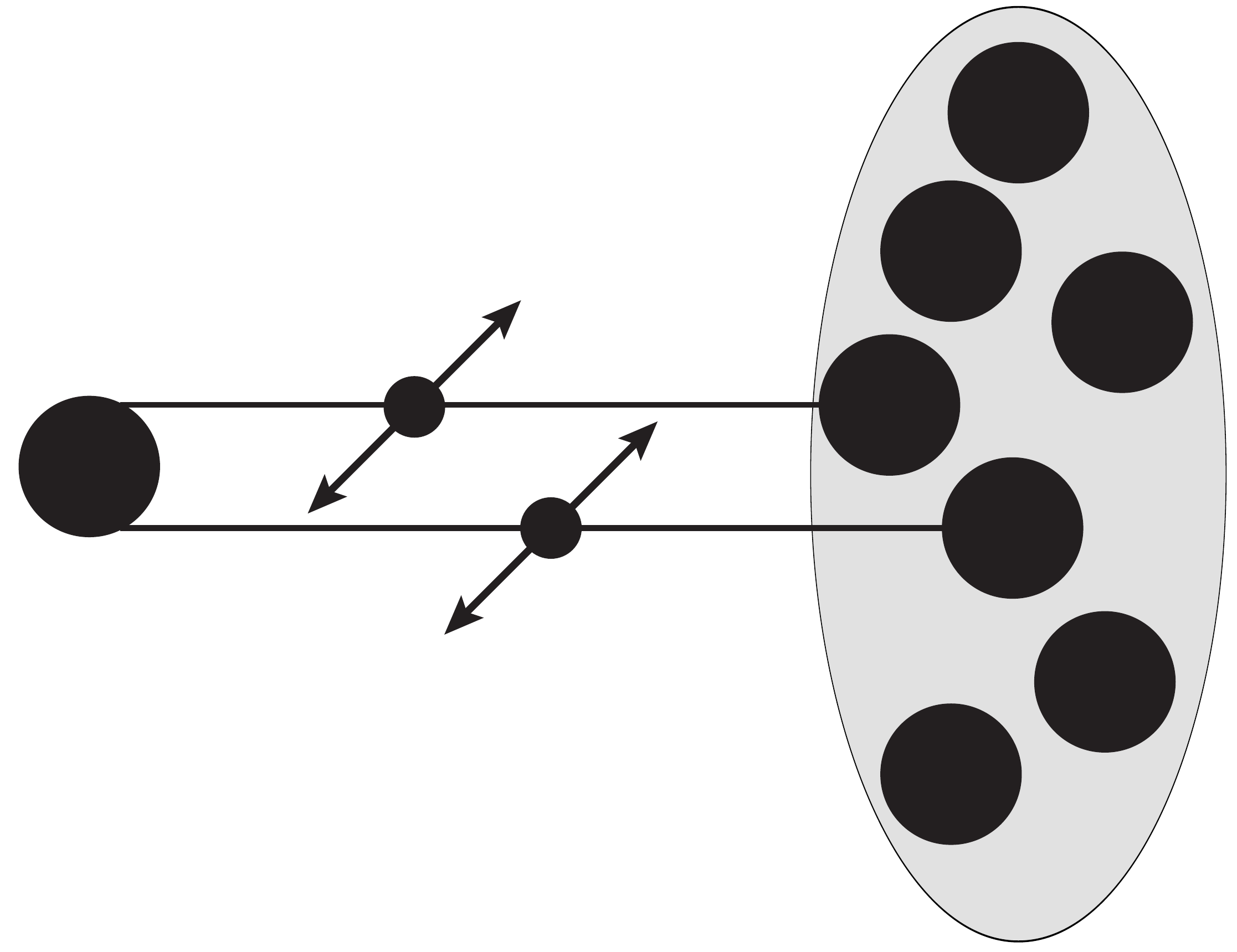}} \\b)
\end{minipage}
\caption{A schematic representation of some possible DPS processes in p$A$ collision: a) DPS occur between one incident proton and one nucleon. 
b) DPS occur between one  incident proton and \textit{two different} nucleons.} 
\label{f:dps_in_pA_contributions}
\end{figure}

The composite nature of a nuclear target leads to various DPS
contributions which are absent in pp (p$\bar{\rm p}$)
collisions. Apart from the ``standard'' DPS process in shown in
Fig.~\ref{f:dps_in_pA_contributions} a), one can have a DPS process
involving one incident proton and two different nucleons, as
schematically shown in Fig.~\ref{f:dps_in_pA_contributions} b). In the
following we will refer to processes in
Fig.~\ref{f:dps_in_pA_contributions} a) and
Fig.~\ref{f:dps_in_pA_contributions} b) as to DPS~I and DPS~II
contributions respectively.

Since DPS~I and DPS~II contributions involve different number of
nucleons it is quite natural to expect a different dependence of the
corresponding total cross sections on an atomic mass number\footnote{To the best of our knowledge, a similar but somewhat different
  assumption was first made by Goebel, Halzen and Scott in
  1980 \cite{Goebel:1979mi}. Namely, it was postulated that total
  cross sections for DPS and ``standard'' \textit{single parton scattering}  processes in p$A$
  collision will have a different $A$-dependence. However, no
  distinction between \mbox{DPS~I} and \mbox{DPS~II} contributions was
  made and corresponding expressions for total cross sections were not
  provided.} $A$. In 2001 Strikman and Treleani have published the pioneering
work \cite{Strikman:2001gz} where expression for the total cross
sections for DPS~I and DPS~II processes were given for the first time.
Within their model a total DPS cross section for p$A$ collisions,
assuming no interference between both DPS processes, can be written as
a sum of two terms
\begin{eqnarray}
	\sigma^{\rm DPS}_{\pA} = \sigma^{\rm DPS}_{\rm I} + \sigma^{\rm DPS}_{\rm II},
	\label{eq:dps_pA_two_terms}
\end{eqnarray}
where $\sigma^{\rm DPS}_{\rm I}$ can be expressed, neglecting a
difference between proton and neutron, in terms of a total DPS cross
section for pp collisions as
\begin{eqnarray}
  \sigma^{\rm DPS}_{\rm I} =  {A} \, \sigma^{\rm DPS}_{\rm pp}
  = \frac{A}{1 + \delta_{ab}}\frac{\sigma_a\sigma_b}{\sigma_{eff}},
  \label{eq:DPSI}
\end{eqnarray}
where $\sigma_{eff}$ is an effective DPS interaction area 
and we use $1 + \delta_{ab}$ in denominator in order to reflect
the fact that one has to divide a total cross section by 2 for
production of two indistinguishable final states. We see that
$\sigma^{\rm DPS}_{\rm I}$ scales simply as a total number of nucleons
$A$. The DPS~II contribution, however, scales differently.  The
expression for $\sigma^{\rm DPS}_{\rm II}$ was found to be equal to
\begin{eqnarray}
  \sigma^{\rm DPS}_{\rm II} &=& 
  \frac{1}{1 + \delta_{ab}}\frac{A - 1}{A}\sigma_a\sigma_b
  \int d^2s \, {T}^2_{A} ({\bf s}) =\nonumber\\
  &=& 
  \frac{1}{1 + \delta_{ab}}\sigma_a\sigma_b \, {F_{\pA}},
\end{eqnarray}
where
\begin{eqnarray}
  {F_{\pA}} = \frac{A - 1}{A}\int d^2s \, {T^2_A}({\bf s})
\end{eqnarray}
and the factor $(A -1)/{A}$ is the number of possible nucleon
pairs $A(A-1)$ divided by $A^2$ which comes from normalization of a two nucleon
form-factor\footnote{The factor $(A - 1) / A$ was absent in the
  original publication of Strikman and Treleani
  \cite{Strikman:2001gz}. It was also absent in a work of Cattaruzza, Del Fabbro
  and Treleani \cite{Cattaruzza:2004qb} which followed
  \cite{Strikman:2001gz}. To the best of our knowledge, this factor appears first
  in the work of Frankfurt \textit{et al.} \cite{Frankfurt:2004kn} and
  a detailed derivation was later given in the work of \mbox{Blok
    \textit{et al.} \cite{Blok:2012jr}}.}  and ${T}_{A}$ is a
nuclear density function $\rho_{A}\left(r\right)$ integrated over a
longitudinal coordinate
\begin{eqnarray}
  {T_A}\left({\bf s}\right) = \int dz \, \rho_{A}\left({\bf s}, z\right),
\end{eqnarray}
where $\rho_{A}\left(r\right)$ obeys a standard normalisation condition
\begin{eqnarray}
	\int d^3r \,\rho_{A}\left(r\right) = {A}.
	\label{eq:normalization_rho}
\end{eqnarray}
Note that only the DPS~I contribution depends on $\sigma_{eff}$ which, in
turn, is sensitive to partonic correlations in a transverse plane
of a hadron, see \cite{Frankfurt:2004kn, Calucci:1997ii, Frankfurt:2003td, Calucci:1999yz, Calucci:2009sv, Calucci:2010wg, Rogers:2009ke, Domdey:2009bg, Flensburg:2011kj, Blok:2013bpa} and
the review \cite{Diehl:2013mma}.

Combining DPS~I and DPS~II contributions together one  can write  Eq.~\eqref{eq:dps_pA_two_terms} as 
\begin{eqnarray}
  \sigma^{\rm DPS}_{\pA} =
  \sigma^{\rm DPS}_{\rm pp}\left({A} + \sigma_{eff} {F_{\pA}}\right).
  \label{eq:dps_pA_correlations}
\end{eqnarray}
We see that within Strikman $\&$ Treleani approach one can express the
difference between $\sigma^{\rm DPS}_{\rm pp}$ and
$\sigma^{\rm DPS}_{\pA}$ solely in terms of a geometrical quantity
${T_A}({\bf s})$ which, in turn, depends on a distribution of
matter in a given nucleus.  In order to perform numerical evaluations
with this formula one has to specify a form of the nuclear matter
density function $\rho_{A}$ which we choose to have a shape of the
Woods-Saxon potential \cite{Woods:1954zz}
\begin{eqnarray}
  \rho_{A}\left(r\right) =
  \rho_0\frac{1 + \omega\left(r / R_{A}\right)^2}%
  {1 + \exp\left[(r - R_{A})/a\right]},
  \label{eq:eq:wood_saxon_rho}
\end{eqnarray}
where $R_{A}$ is a nuclear radius, $a$ is a length of smearing of
a nuclear surface, $\omega$ describes a deviation from a spherical
form and a value of $\rho_0$ is fixed by
Eq.~\eqref{eq:normalization_rho}.  If one consider a spherical nucleus the
Woods-Saxon nuclear matter density function reduces to the Fermi
distribution
\begin{eqnarray}
  \rho_{A}\left(r\right) =
  \frac{\rho_0}{1 + \exp\left[(r - R_{A})/a\right]}.
  \label{eq:eq:fermi_rho}
\end{eqnarray}
In order to perform numerical evaluations we need to choose a special
parametrisation of a nuclear matter density.  In this work we use a
parametrisation of the GLISSANDO 2 code \cite{Rybczynski:2013yba} (the
same parametrisation as in \pytppp). Namely, for nuclei with mass
numbers in a range $4 \le {A} \le 208$ we use Wood-Saxon (Fermi)
profile given by Eq.~\eqref{eq:eq:fermi_rho} with
\begin{eqnarray}
  R_{A} &=& \left[1.10 {A}^{1/3} - 0.656 {A}^{-1/3}\right]\,{\rm fm},\\
  a &=& 0.459 \,{\rm fm}, 
\end{eqnarray}
which corresponds to spherical nuclei with a nucleon-nucleon ($NN$)
repulsion distance equal to $d = 0.9 \,{\rm fm}$.

\begin{widetext}
\begin{figure}
\center{\includegraphics[width=1.0\linewidth]{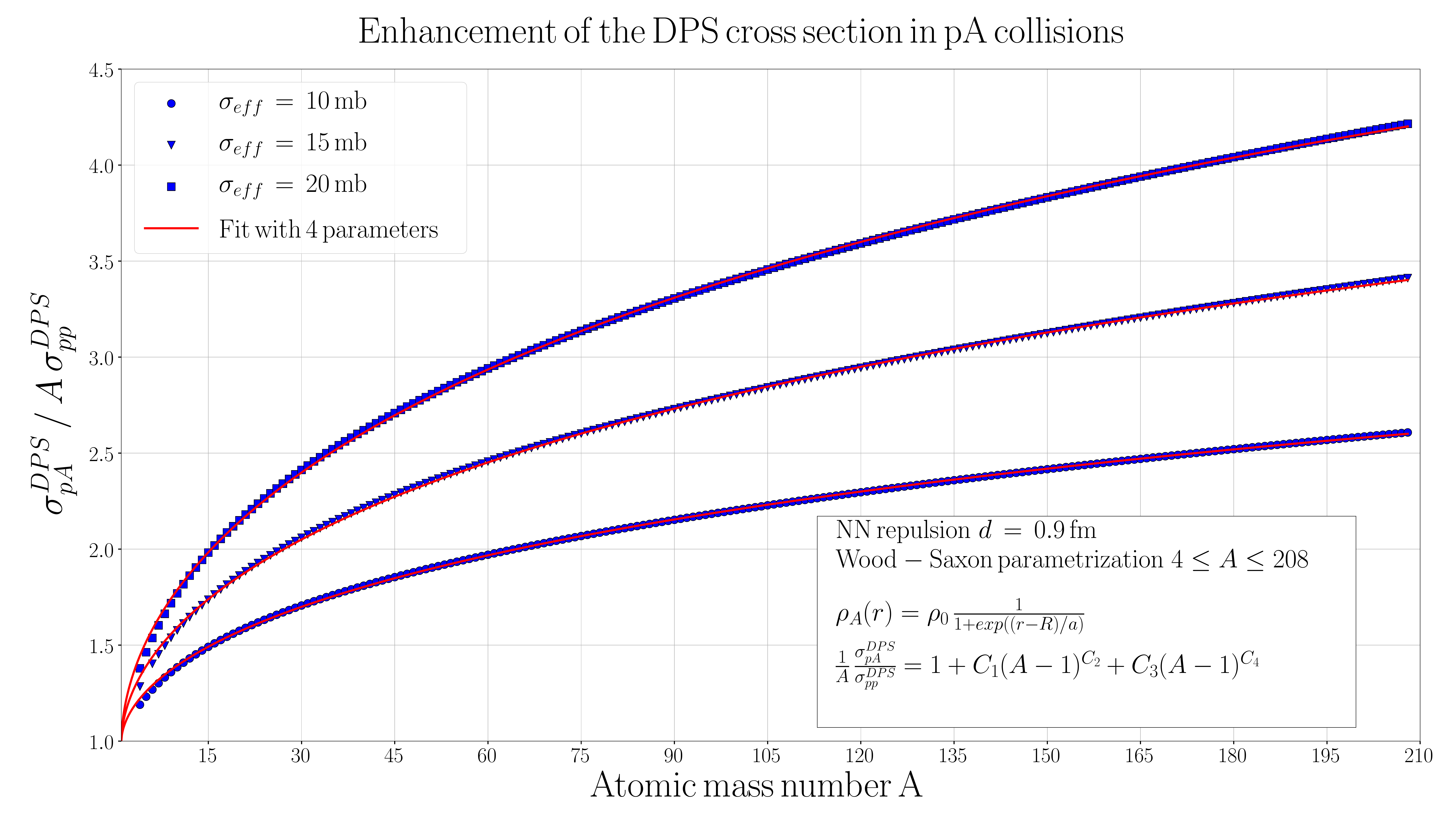}}
\caption{Enhancement of the $\sigma_{\pA}^{\rm DPS}$ with respect
  to $\sigma_{\rm pp}^{\rm DPS}$ normalised according to the atomic
  mass number A.  Wood-Saxon (Fermi) form of the nuclear matter
  distribution $\rho_{A}(r)$ with parameters taken from
  \cite{Rybczynski:2013yba}.}
\label{fig:A_enhancement}
\end{figure}  
\end{widetext}

Now we can evaluate ${A} + \sigma_{eff} {F_{\pA}}$. In order to do
that we vary $\sigma_{eff}$ in between $10 \, {\rm mb}$ and
$20 \, {\rm mb}$ which agrees with most experimental studies of
four-jet DPS production. In Fig.~\ref{fig:A_enhancement} we plot the ratio
$\sigma^{\rm DPS}_{\pA} / {A} \, \sigma^{\rm DPS}_{\rm pp} = 1
+ \frac{1}{A} \, \sigma_{eff} {F_{\pA}}$ as a function of
$A$.  In the absence of the second term in
Eq.~\eqref{eq:dps_pA_two_terms} this ratio would always be equal to
unity. However, we see that a total DPS cross section for heavy nuclei
in pA collisions is about $3A$ times bigger as a
corresponding one in pp collisions. We also see that variation of
$\sigma_{eff}$ leads to significant changes in behaviour of
$\sigma^{\rm DPS}_{\pA} / {A} \, \sigma^{\rm DPS}_{\rm
  pp}$. Such numerical estimate were first made in
\cite{Strikman:2001gz} and the enhancement $\sim 3A$ was later
given in \cite{Frankfurt:2004kn, Blok:2012jr, Cattaruzza:2004qb, dEnterria:2012jam}.
It is handy to approximate a behaviour of the DPS enhancement factor
$\sigma^{\rm DPS}_{\pA} / {A} \, \sigma^{\rm DPS}_{\rm pp}$ as
\begin{eqnarray}
  \frac{1}{A} \, \frac{\sigma^{\rm DPS}_{\pA}}{\sigma^{\rm DPS}_{\rm pp}} =
  1 + C_1 ({A} - 1)^{C_2} +  C_3 ({A} - 1)^{C_4},
\label{eq:fitting_function}
\end{eqnarray} 
where a second term was added to correctly describe enhancement for
heavy nuclei and the coefficients $C_1$ - $C_4$ can be identified by 
fitting Eq.~\eqref{eq:fitting_function} to our simulations, as it is
shown in Fig.~\ref{fig:A_enhancement}. This fitting may look somewhat superfluous, since computations within the
\mbox{Strikman $\&$ Treleani}  framework  are not time consuming, 
however, its advantage will become clear later when we will discuss our Monte Carlo simulations.

It is important to note that in general two terms in
Eq.~\eqref{eq:dps_pA_two_terms} correspond to different phase spaces and
thus a factorised form of a total DPS cross section given in
Eq.~\eqref{eq:dps_pA_correlations} violates conservation of a
longitudinal momentum. It becomes clear if one writes down expressions for $\sigma^{\rm DPS}_{\rm I}$ and $\sigma^{\rm DPS}_{\rm II}$
\begin{widetext}
\begin{eqnarray}
	\sigma^{\rm DPS}_{\rm I} &=&
	\frac{{A}}{\sigma_{eff}}
	\sum\limits_{a_1, a_2, b_1, b_2}
	\int \, \prod\limits^4_{i = 1} dx_i \,
	D_{a_1, a_2}(x_1, x_2) \,
	D_{b_1, b_2}(x_3, x_4) \,
	\hat{\sigma}_{a_1, b_1} \, 
	\hat{\sigma}_{a_2, b_2},
	\label{eq:strikman_treleani_dPDFs_I}\\
	\sigma^{\rm DPS}_{\rm II} &=&
	{F_{\pA}}	
	\sum\limits_{a_1, a_2, b_1, b_2}
	\int \, \prod\limits^4_{i = 1} dx_i  \,
	D_{a_1, a_2}(x_1, x_2) \,
	f_{b_1}(x_3) \,
	f_{b_2}(x_4) \, 
	\hat{\sigma}_{a_1, b_1} \, 
	\hat{\sigma}_{a_2, b_2},
	\label{eq:strikman_treleani_dPDFs_II}
\end{eqnarray}
\end{widetext}
where $f_{i}(x_i)$ are standard collinear PDFs and $D_{i,j}(x_1, x_2)$
are so called \textit{double parton distribution functions} which give
a probability to find two partons of flavour $i,j$ with Bjorken-$x$'es
$x_1,x_2$ in a proton\footnote{Here we have omitted dependence on factorisation and renormalisation scales.}. Assuming no correlations in $x$-space,
Eq.~\eqref{eq:strikman_treleani_dPDFs_I} and
Eq.~\eqref{eq:strikman_treleani_dPDFs_II} can be written as
\begin{widetext}
\begin{eqnarray}
	\sigma^{\rm DPS}_{\rm I}  &=&
	\frac{A}{\sigma_{eff}}
	\sum\limits_{a_1, a_2, b_1, b_2}
	\int \, \prod\limits^4_{i = 1} dx_i \,
	f_{a_1}(x_1) \,
	f_{a_2}(x_2) \,
	f_{b_1}(x_3) \,
	f_{b_2}(x_4) \, 
	\theta(1 - x_1 - x_2) \,
	\theta(1 - x_3 - x_4) \,
	\hat{\sigma}_{a_1, b_1} \, 
	\hat{\sigma}_{a_2, b_2},
    \label{eq:dps_pA_exact_phase_space_1}\\
	\sigma^{\rm DPS}_{\rm II}  &=&
	{F_{{\rm p}A}}	
	\sum\limits_{a_1, a_2, b_1, b_2}
	\int \, \prod\limits^4_{i = 1} dx_i \,
	f_{a_1}(x_1) \,
	f_{a_2}(x_2) \,
	f_{b_1}(x_3) \,
	f_{b_2}(x_4) \, 
	\theta(1 - x_1 - x_2) \,
	\theta(x_3) \,
	\theta(x_4) \,
	\hat{\sigma}_{a_1, b_1} \, 
	\hat{\sigma}_{a_2, b_2}.
	\label{eq:dps_pA_exact_phase_space_2}
\end{eqnarray}
\end{widetext}
We see that different constraints on Bjorken-$x$'es lead to different
integration regions which does not let us to write a total DPS cross
section as in Eq.~\eqref{eq:dps_pA_correlations}. However, the
difference between two phase spaces should become relevant only for
large $x$'es where dPDFs have relatively small values and therefore
their impact on a total DPS cross section is small. 
The direct numerical check gives the difference which is well below the percent level and, therefore, is completely negligible.

\section{\label{s:angantyr_model} Angantyr and MPI models}
Now let us turn our attention to a Monte Carlo approach to p$A$
collisions. Usually in this field existing Monte Carlo event
generators are more ``special purpose'' and mostly dedicated to
studies of formation and evolution of the \textit{Quark-Gluon Plasma},
\textit{e.g.} \mbox{EPOS-LHC} \cite{Pierog:2013ria}, AMPT
\cite{Lin:2004en} and HIJING \cite{Wang:1991hta}. From the other side
there are models postulating flow-like effects to have a non-thermal
origin and therefore aiming to reproduce general features of p$A$ (AA)
collisions by adding a nuclear structure ``on top'' of existing pp
models. One such model is called \angantyr \cite{Bierlich:2018xfw} was
recently implemented into \pytppp event generator. It was inspired by
the old Lund Fritiof model \cite{Andersson:1986gw} and the DIPSY model\footnote{DIPSY models
  BFKL evolution of a gluon cascade via Mueller dipole approach
  \cite{Mueller:1993rr}, \cite{Mueller:1994jq}}
\cite{Avsar:2006jy, Avsar:2005iz, Flensburg:2011kk}.
\begin{figure}
\centering
\begin{minipage}[h]{0.44\linewidth}
\center{\includegraphics[width=0.9\linewidth]{pA_type_1.pdf}} \\a)
\end{minipage}
\hfill
\begin{minipage}[h]{0.44\linewidth}
\center{\includegraphics[width=0.9\linewidth]{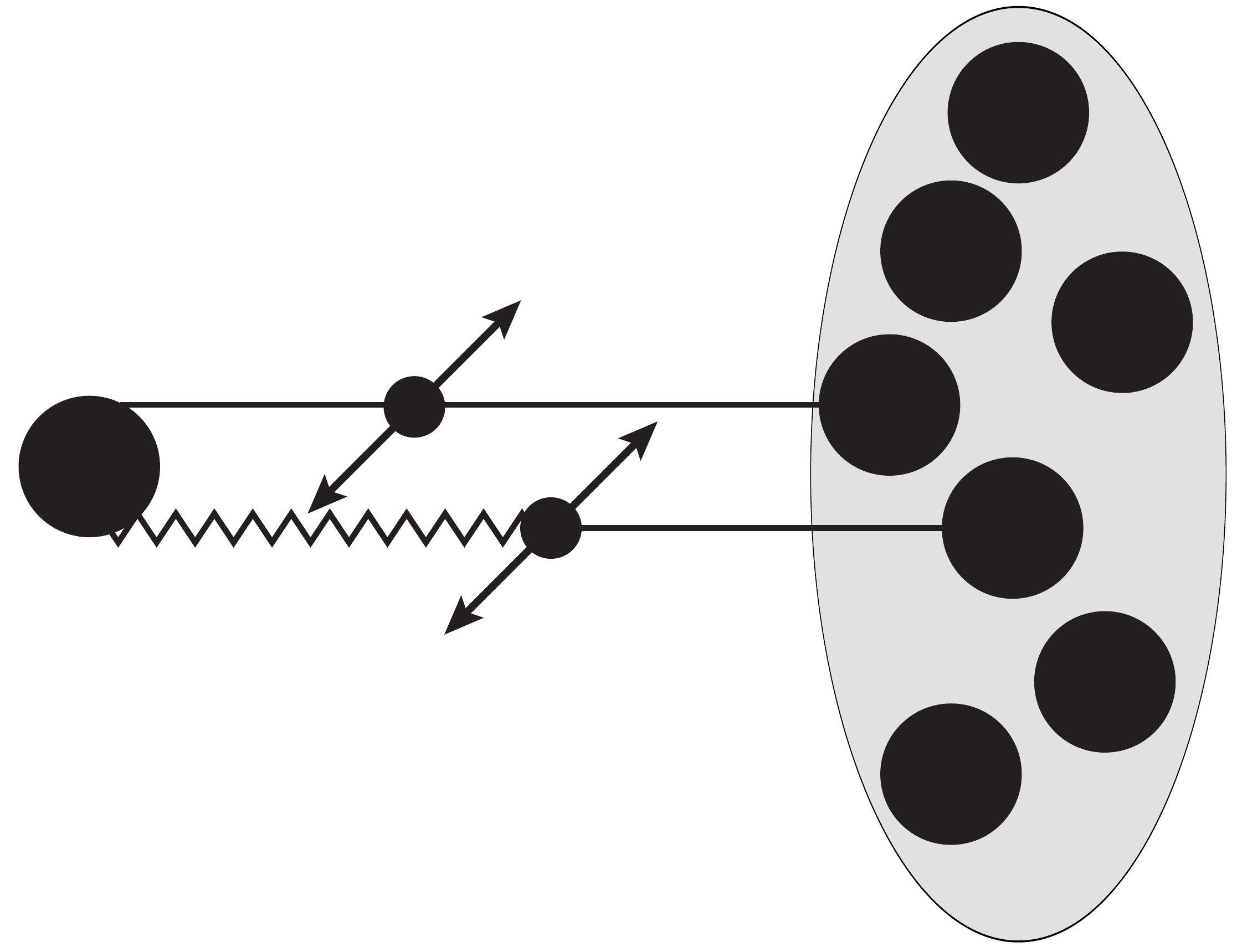}} \\b)
\end{minipage}
\cprotect\caption{A schematic representation of DPS processes in p$A$
  collision according to the \angantyr model: a) DPS occur between one
  incident proton and one nucleon (modelled with a standard MPI
  machinery).  b) DPS occur between one incident proton and
  \textit{two different} nucleons (modelled with a simplified MPI
  machinery and modified diffractive machinery).  A zigzag line here
  corresponds to a Pomeron inside of an incident proton.}
\label{f:dps_in_pA_contributions_Angantyr}
\end{figure}

The production of final state particles in \angantyr is based upon
\pythia's models for \textit{multiple parton interaction} (MPI)
\cite{Sjostrand:1985vv, Sjostrand:1987su, Sjostrand:2004pf, Corke:2009tk, Corke:2011yy} and diffractive
processes \cite{Navin:2010kk}, \cite{Corke:2010yf} with certain
modifications which will be explained below.

First of all let us describe the way \pythia treats production of
particles in interactions involving one incident proton and a single
nucleon as in Fig.~\ref{f:dps_in_pA_contributions_Angantyr} a). In
this case one could naively expect that all MPIs would be distributed
according to a Poissonian distribution. This approach to MPI
modelling, however, may lead to momentum violation and is in
contradiction with KNO scaling \cite{Koba:1972ng} of charged
multiplicity distributions, see review \cite{Sjostrand:2017cdm}. In
order to solve this issues all MPIs in \pythia are ordered in
transverse momentum as
$\sqrt{s}/2 > p_{\perp 1} > p_{\perp 2} > \ldots > p_{\perp n} >
p_{\perp {\rm min}}$.  A probability of a first interaction to happen
at a given transverse momentum
$d\sigma \, / \, dp_{\perp1}  \, \sigma^{\rm pp}_{\rm ND}\left(s\right)$
is multiplied by a Sudakov-like exponent
\begin{eqnarray}
  \frac{d\mathcal{P}}{dp_{\perp1}} &=& 
  \frac{1}{\sigma^{\rm pp}_{\rm ND}\left(s\right)} \frac{d\sigma}{dp_{\perp1}}
  \times\nonumber\\
  &\times&
  \exp\left(
  -\int\limits^{\sqrt{s}/2}_{p_{\perp1}} 
  \frac{1}{\sigma^{\rm pp}_{\rm ND}\left(s\right)} 
  \frac{d\sigma}{dp^\prime_\perp} dp^\prime_\perp
  \right), 
\end{eqnarray}
which ensures that no other interactions will happen in 
$p_\perp$-range between $\sqrt{s} / 2$ and $p_{\perp 1}$. Therefore, a
probability for all subsequent interactions is given by
\begin{eqnarray}
  \frac{d\mathcal{P}}{dp_{\perp i}} &=& 
  \frac{1}{\sigma^{\rm pp}_{\rm ND}\left(s\right)} \frac{d\sigma}{dp_{\perp i}}
  \times\nonumber\\
  &\times&
  \exp\left(
  -\int\limits^{p_{\perp i-1}}_{p_{\perp i}} 
  \frac{1}{\sigma^{\rm pp}_{\rm ND}\left(s\right)} 
  \frac{d\sigma}{dp^\prime_\perp} dp^\prime_\perp
  \right),
\end{eqnarray}
which ensures the $p_\perp$ ordering. In addition to it, the MPI model
of \pythia accounts for momentum and number conservation which implies
that PDFs used for a second interaction (as well as for all subsequent
interactions) will be ``squeezed'' and reweighted according to a
history of all previous interactions in order to take into account
changes in the parton content and preserve momentum conservation, see
\cite{Sjostrand:2004pf}. It should be
noted that $\sigma_{eff}$ does not enter explicitly into this
model. More specifically a ratio
$\sigma^{\rm pp}_{\rm ND}\left(s\right) / \sigma_{eff}$ describes a
deviation of a distribution of MPIs from a Poissonian distribution,
see \cite{Seymour:2013sya}, \cite{Sjostrand:2004pf}, and review
\cite{Sjostrand:2017cdm}. However it implies that direct comparison
between predictions of MPI model of \pythia and DPS model of Strikman
and Treleani is not possible meaning that a corresponding value of
parameter $\sigma_{eff}$ is unknown. We will come back to this issue
later in Section~\ref{s:predictions_of_pythia}.

Description of processes involving one incoming proton and \textit{two
  different} nucleons is somewhat more complicated. In principle one
should have implemented the same contribution as it is shown in
Fig.~\ref{f:dps_in_pA_contributions} b).  However, in practice,
incorporation of such processes into \pythia's framework leads to
serious technical difficulties. It is possible to circumvent these
issues by mimicking a second absorptive interaction as in
Fig.~\ref{f:dps_in_pA_contributions} b) via nucleon-Pomeron collision
as in \mbox{Fig.~\ref{f:dps_in_pA_contributions_Angantyr}
  b)}. Therefore, in order to simulate double absorptive process from
Fig.~\ref{f:dps_in_pA_contributions} b), \angantyr will first simulate
a single absorptive process via a standard pp machinery and then
simulate a second absorptive process \textit{as if} was produced
through a single diffractive excitation, much in the spirit of the old
Fritiof model. All subsequent interactions will be produced via
standard pp or proton-Pomeron MPI machinery. Energy-momentum
conservation is ensured when extracting the fictitious Promeron
from the projectile proton, but this will not influence the PDFs of
the proton, and except for the proton remnants the primary absorptive
process will look exactly like a normal non-diffractive pp
event.

There are several ways to produce diffractive events in \pytppp. The
\angantyr model is based upon a model of \textit{soft} diffraction of
\pythia. For high-mass diffraction \pythia uses the Ingelman and
Schlein model where the Pomeron is treated as a hadronic state
\cite{Ingelman:1984ns}. Within this approach \pythia treats a
proton-Pomeron collision as a normal non-diffractive hadron-hadron
collision with standard MPI, initial and final state radiation
machinery. Therefore, a corresponding differential $2\rightarrow2$
cross section is given by
\begin{eqnarray}
  d\sigma^{\rm p\Pom}_{ij}  &=& \frac{dx_\Pom}{x_\Pom} \,
  \frac{dx_1}{x_1} \, \frac{d\beta}{\beta} 
  \times\nonumber\\
  &\times&
  F\left(x_\Pom\right) \, x_1 f_{i}(x_1, Q^2) \,
  \beta f_{j/\Pom}(\beta, Q^2) \,
  d\hat{\sigma}_{ij},
  \label{eq:pythia_ingelman}
\end{eqnarray}
where $x_\Pom$ is a fraction of the target proton momentum taken by
the Pomeron, $\beta$ is a fraction of the Pomeron's momentum taken by
the parton $j$ and $x_1$ is a fraction of Pomeron's momentum taken by
parton $i$. A diffractive mass $M^2_X$ is therefore given by
$M^2_X = x_\Pom s$. In the \angantyr model a Pomeron flux
$F\left(x_\Pom\right)$ is by default taken to be a constant which
implies a flat distribution in $\log\left(M^2_X\right)$, although this
can be changed in the settings.  The hard cross section
$\hat{\sigma}_{ij}$ in Eq.~\eqref{eq:pythia_ingelman} is the standard
\textit{leading order} (LO) $2\rightarrow2$ cross section which is
known to be divergent for low $p_\perp$ values. As in the pp case
\pythia imposes a smooth cut-off on $\hat{\sigma}_{ij}$ according to
\begin{eqnarray}
  \frac{d\hat{\sigma}_{ij}}{dp^2_\perp} \varpropto
  \frac{\alpha^2_s(p^2_\perp)}{p^4_\perp} \rightarrow 
  \frac{\alpha^2_s(p^2_\perp + p^2_{\perp0})}{(p^2_\perp + p^2_{\perp0})^2},
  \label{eq:pythia_cut_off}
\end{eqnarray}
where $p_{\perp0}$ is a soft regulator which depends either on
diffractive mass (for diffractive processes) or on collision energy
(for standard pp processes).  Nevertheless, even after a
regularisation of $\hat{\sigma}_{ij}$ as in
Eq.~\eqref{eq:pythia_cut_off}, an integrated partonic cross section may
exeed a total non-diffractive proton-Pomeron cross section for a given
diffractive mass $M_X$. In the MPI model of \pythia it is interpreted
as a possibility to have several sub-scatterings in each collision
with an average number
\begin{widetext}
\begin{eqnarray}
  \langle N^{\rm p\Pom}_{sc}\left(M^2_X\right) \rangle =
  \frac{1}{\sigma^{\rm p\Pom}_{\rm ND}\left(M_X\right)} \,
  \int \frac{dx_1}{x_1} \, \frac{d\beta}{\beta} \, dp^2_\perp \,
  \sum\limits_{ij} x_1 f_{i}(x_1, Q^2) \, \beta f_{j/\Pom}(\beta, Q^2) \,
  \frac{d\hat{\sigma}_{ij}}{dp^2_\perp}.	
  \label{eq:n_sub_col_SD}					   
\end{eqnarray}
\end{widetext}
However, as it was pointed out in \cite{Bierlich:2016smv}, a modelling
of single absorptive events via \textit{single diffractive} (SD)
events results in too low activity in p$A$ collision.  In principle, one
can solve this problem either by tuning the value of
$\sigma^{\rm p\Pom}_{\rm ND}\left(M_X\right)$ in
Eq.~\eqref{eq:n_sub_col_SD} or by changing Pomeron PDFs. By comparing a
distribution $d\langle N^{\rm p\Pom}_{sc}\rangle / dy$ for SD events
\begin{widetext}
\begin{eqnarray}
  \frac{d\langle N^{\rm p\Pom}_{sc}\rangle}{dy} =   
      \frac{1}{\sigma^{\rm p\Pom}_{\rm ND}\left(M^2_X\right)} \,
      \int \frac{dx_1}{x_1} \, \frac{d\beta}{\beta} \, dp^2_\perp \,
      \sum\limits_{ij} x_1 f_{i}(x_1, Q^2) \, \beta f_{j/\Pom}(\beta, Q^2) \,
      \frac{d\hat{\sigma}_{ij}}{dp^2_\perp}
      \delta\left(y - \frac{1}{2}\log\frac{x_1}{\beta x_\Pom}\right),
  \label{eq:n_sub_col_SD_at_y}														   
\end{eqnarray}
\end{widetext}
against a corresponding distribution for standard non-diffractive pp events
\begin{widetext}
\begin{eqnarray}
  \frac{d\langle N^{\rm pp}_{sc}\rangle}{dy} =  
      \frac{1}{\sigma^{\rm pp}_{\rm ND}\left(s\right)} \,
      \int \frac{dx_1}{x_1} \, \frac{dx_2}{x_2} \, dp^2_\perp \,
      \sum\limits_{ij} x_1 f_{i}(x_1, Q^2) \, x_2 f_{j}(x_2, Q^2) \, 
      \frac{d\hat{\sigma}_{ij}}{dp^2_\perp}
      \delta\left(y - \frac{1}{2}\log\frac{x_1}{x_2}\right),
  \label{eq:n_sub_col_pp}
\end{eqnarray}
\end{widetext}
we see that if in Eq.~\eqref{eq:n_sub_col_SD_at_y} we set
$\beta f_{j/\Pom}(\beta, Q^2) \rightarrow x_\Pom \beta
f_{j}\left(x_\Pom\beta, Q^2\right)$,
$\sigma^{\rm p\Pom}_{\rm ND}\left(M_X\right) \rightarrow \sigma^{\rm
  pp}_{\rm ND}\left(s\right)$ then we get an expression very similar
to Eq.~\eqref{eq:n_sub_col_pp}. Also if the energy dependence of soft
regularisation in $\hat{\sigma}$ is changed from
$p_{\perp0}\left(M^2_X\right)$ to $p_{\perp0}\left(s\right)$, the
expression will be identical for large negative rapidities, which is
what is desired.

The validity of this approach was studied in detail in
\cite{Bierlich:2018xfw}. In particular it was shown that
Eq.~\eqref{eq:n_sub_col_SD_at_y} modified as described, provides an
overall fair description of experimental data. However, all Angantyr checks in
\cite{Bierlich:2018xfw} were related to MPI-sensitive distributions
like, for example, a charged multiplicity distribution. Indeed, such
distributions are known to be very sensitive to a number of semi-hard
and soft sub-collisions in a given event, see, for example, review
\cite{Sjostrand:2017cdm}. Therefore, correct predictions for a shape
of such distributions can be seen as a validation of both MPI and
\angantyr models. In the next section of this paper we will switch our
attention from MPI to DPS processes and perform another check of the
\angantyr model.  Namely, we will study how well it can reproduce
predictions of Strikman and Treleani for DPS production of four
\textit{hard} jets in p$A$ collisions.

\section{\label{s:predictions_of_pythia} Predictions of \protect\pythia}
Before starting to compare predictions of \pythia against Strikman
$\&$ Treleani model several important comments have to be made.  First
of all, as we already mentioned in Section~\ref{s:angantyr_model}, all
MPIs produced in a given event are strictly ordered in $p_\perp$. This
ordering may seem to be in contradiction with the Strikman $\&$
Treleani model, where the two processes are treated equal. However,
here we will only consider the case of having two identical processes,
so this is then just a trivial numbering
issue. One should also keep in mind that, in order to derive
Eq.~\eqref{eq:dps_pA_correlations}, Strikman and Treleani neglected
partonic correlations in $x$-space and assumed that both DPS~I and
DPS~II contributions populate the same phase space region. As we have
noticed in Section~\ref{s:strikman_treleani_model}, the error due to
this approximation is completely negligible.  Effects due to the
correlations in $x$-space, nevertheless, may have a sizeable impact,
see \cite{Blok:2012jr} and \mbox{\cite{Korotkikh:2004bz, Cattaruzza:2005nu, Gaunt:2009re}}.  We also should keep in mind
that \pythia's approach to momentum and number conservation
effectively means presence of non-trivial $x$-space partonic
correlations in the MPI machinery. Finally, we need to stress that the
parameter $\sigma_{eff}$ does not enter explicitly into the \angantyr
model and, therefore, in order to compare the predictions of Strikman
$\&$ Treleani model against predictions of \angantyr one has to find
the value of $\sigma_{eff}$ in Strikman $\&$ Treleani model by fitting
its predictions to the prediction of \angantyr.

Now, after describing all the important differences between both
approaches, let us study how the DPS enhancement factor
$\sigma^{\rm DPS}_{\pA} / {A} \, \sigma^{\rm DPS}_{\rm pp}$ in
the \angantyr model depends on a total number of nucleons $A$.

Due to the lack of triggering in the MPI machinery one will need to
perform a high number of generation calls in order to collect a good
statistics for a four-jet DPS production, since a second MPI will most
of the time occur at too low scale to be considered as a hard
interaction\cprotect\footnote{\pytppp allows to generate
  \textit{always} two hard interactions in a given event by setting
  \verb|SecondHard:generate = on|. However, usage of this flag
  together with \angantyr is not supported and will lead to wrong
  results.}.  Therefore, we evaluate $\sigma^{\rm DPS}_{\pA}$
according to a following algorithm:
\begin{itemize}
\item Find a total weight $w_{\pA}^{\rm tot}$ for all events
  produced in p$A$ collisions and a corresponding total cross section
  $\sigma^{\rm tot}_{\pA}$.
\item Find a total weight $w_{\pA}^{\rm DPS}$ of all events which
  satisfy a given set of cuts.
\item Find a total DPS cross section in p$A$ collisions
  $\sigma^{\rm DPS}_{\pA}$ from the ratio
  \begin{eqnarray}
    \frac{\sigma^{\rm DPS}_{\pA}}{\sigma^{\rm tot}_{\pA}} =
    \frac{w_{\pA}^{\rm DPS}}{w_{\pA}^{\rm tot}}.
  \end{eqnarray}
\item Repeat the same for pp collisions. Find a corresponding total
  DPS cross section $\sigma^{\rm DPS}_{\rm pp}$.
\item Evaluate
  $\sigma^{\rm DPS}_{\pA} / {A} \sigma^{\rm DPS}_{\rm pp}$.
\end{itemize}

In principle, p$A$ machinery of \pythia allows a user to implement any
isotop with given values of $Z$ and $N$.  Eight nuclei:
$\prescript{4}{}{{\rm He}}$, $\prescript{6}{}{{\rm Li}}$,
$\prescript{12}{}{{\rm C}}$ ,$\prescript{16}{}{{\rm O}}$,
$\prescript{63}{}{{\rm Cu}}$, $\prescript{129}{}{{\rm Xe}}$,
$\prescript{197}{}{{\rm Au}}$ and $\prescript{208}{}{{\rm Pb}}$ are
available by default.  Since a computation of a total DPS cross
section according to the algorithm above can take tens of
hours (depending on a chosen nucleus and a system performance), we
decided to work only with already implemented nuclei and use a fit as
in Eq.~\eqref{eq:fitting_function} for better visualisation of our
results and for comparison against \mbox{Strikman $\&$ Treleani}
model.

Our results for
$\sigma^{\rm DPS}_{\pA} / {A} \sigma^{\rm DPS}_{\rm pp}$ are given in
Tab.~\ref{tab:heavy_nuclei_dist_2}. In our simulations we were
triggering on events with \textit{at least} four jets with
$p_\perp > 20$~GeV. We have also performed a stability check by
varying a parameter \verb|Angantyr:SDTries| controlling the maximum
number of attempts allowed to add a secondary absorptive sub-event (as
in Fig.~\ref{f:dps_in_pA_contributions_Angantyr} b) without violating
energy--momentum conservation. By comparing values
of $\sigma^{\rm DPS}_{\pA} / {A} \sigma^{\rm DPS}_{\rm pp}$ evaluated
at different values of \verb|SDTries| parameter, we see that
fluctuations of
$\sigma^{\rm DPS}_{\pA} / {A} \sigma^{\rm DPS}_{\rm pp}$ do not exceed
a few percent level.

\begin{figure}
\begin{minipage}[h]{1.0\linewidth}
\center{\includegraphics[width=1.0\linewidth]{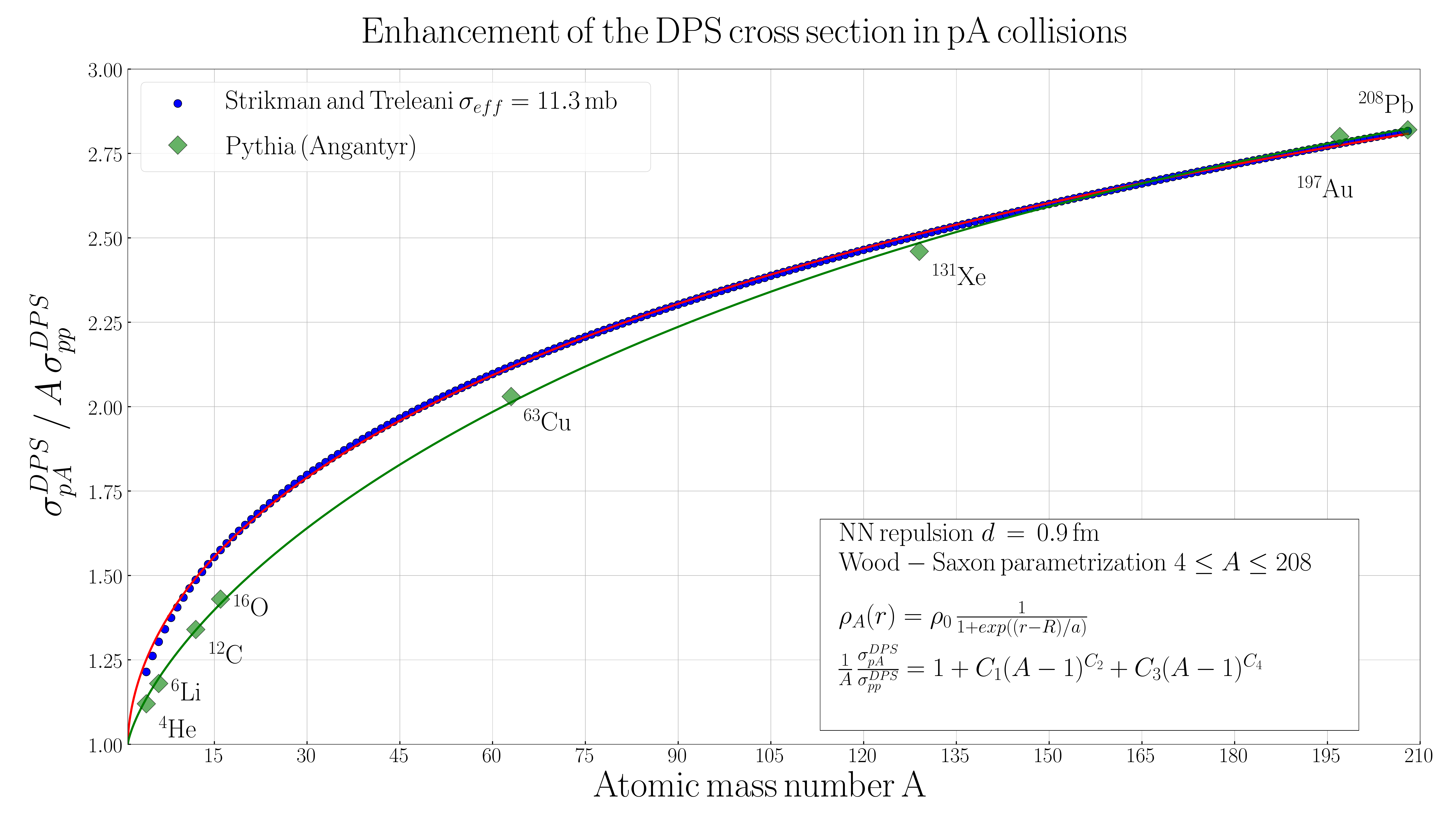}}
\end{minipage}
\cprotect\caption{The DPS enhancement factor
  $\sigma^{\rm DPS}_{\pA} / {A} \, \sigma^{\rm DPS}_{\rm
    pp}$ as a function of a total number of nucleons $A$. Comparison between theoretical predictions of Strikman and
  Treleani \cite{Strikman:2001gz} and \pythia's (\angantyr)
  simulations.}
\label{fig:dps_Angantyr_vs_theory}
\end{figure}

A comparison against Strikman $\&$ Treleani model is given in
Fig.~\ref{fig:dps_Angantyr_vs_theory} and
Tab.~\ref{tab:angantyr_vs_theory}. The \pythia set-up we have used is
given in Appendix~\ref{appendix:pythia_settings}. In order to compare
our results against \mbox{Strikman $\&$ Treleani} model we have tuned
$\sigma_{eff}$ in order to get an agreement in the value of the DPS
enhancement factor
$\sigma^{\rm DPS}_{\pA} / {A} \sigma^{\rm DPS}_{\rm pp}$ for
$\prescript{208}{}{{\rm Pb}}$. We see that by choosing
$\sigma_{eff} = 11.3 \, {\rm mb}$ we can get a satisfactory agreement
between both models for heavy isotopes $\prescript{129}{}{{\rm Xe}}$,
$\prescript{197}{}{{\rm Au}}$ and $\prescript{208}{}{{\rm Pb}}$.

\begin{figure}
\centering
\begin{minipage}[h]{0.44\linewidth}
\center{\includegraphics[width=0.9\linewidth]{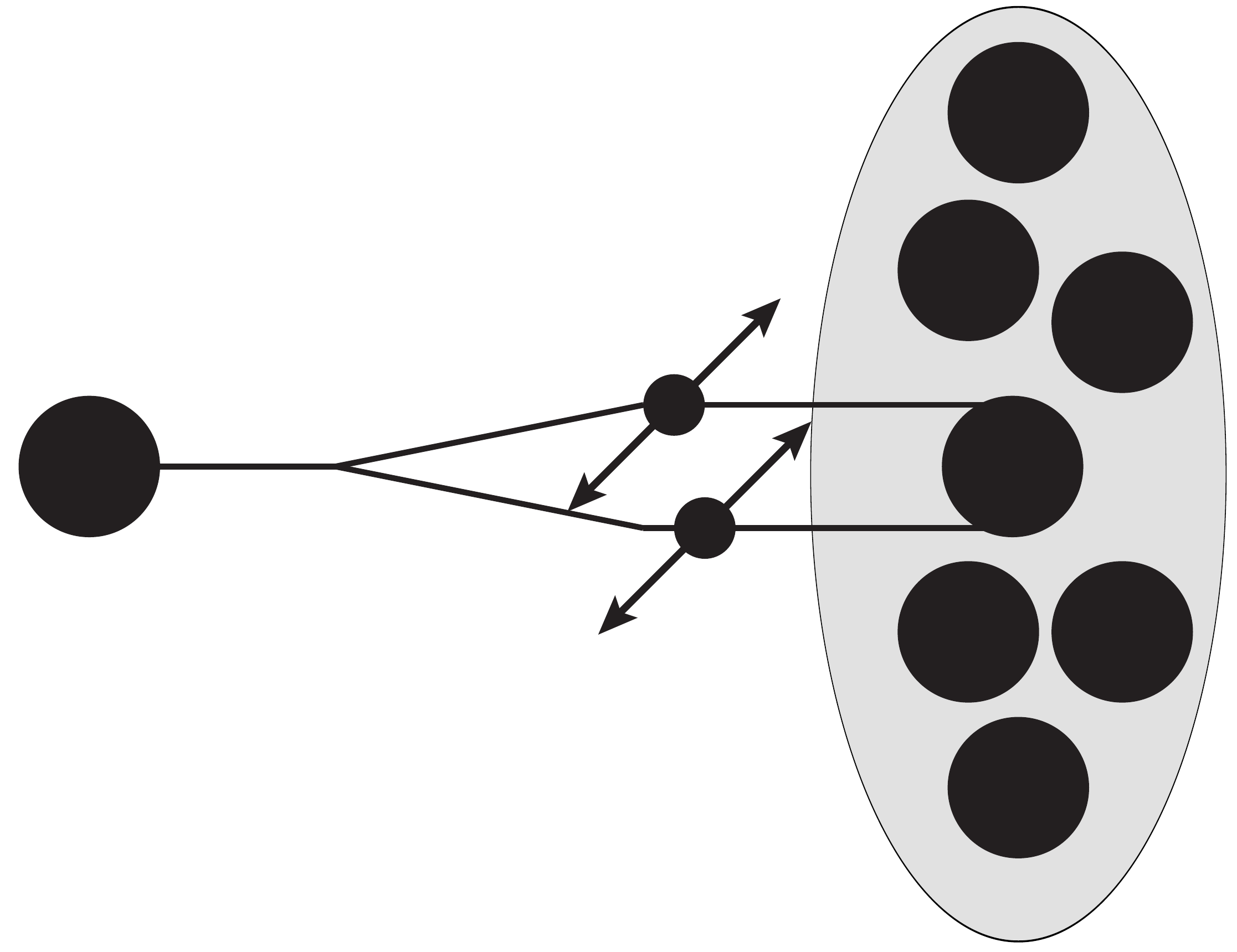}} \\a)
\end{minipage}
\hfill
\begin{minipage}[h]{0.44\linewidth}
\center{\includegraphics[width=0.9\linewidth]{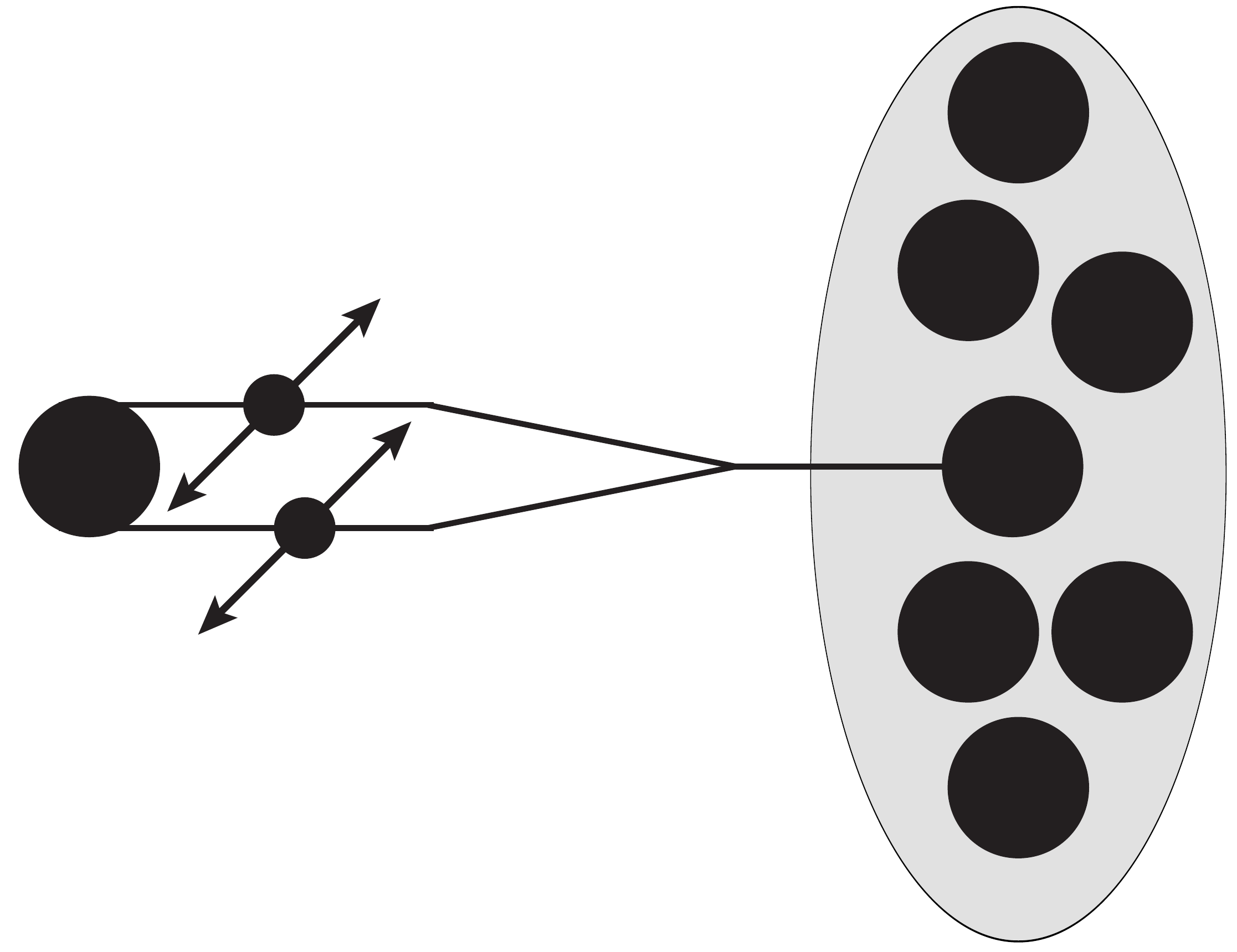}} \\b)
\end{minipage}
\cprotect\caption{A schematic representation of some possible ``$1v2$''
  DPS processes in p$A$ collision: a) A ``$1v2$'' splitting occur in the
  incident proton before two hard interactions take place. b) A
  ``$1v2$'' splitting occur in a nucleon before two hard interactions
  take place. }
\label{f:dps_in_pA_1v2}
\end{figure}

It could be tempting to interpret our simulations as a fake data
and to use Eq.~\eqref{eq:dps_pA_correlations} for a fitting procedure
to extract a value of $\sigma_{eff}$ out of it. However, due to the
differences between the models, such an interpretation would not be
very relevant.
For example, as it was shown in \cite{Blok:2012jr}, perturbative 
splittings of initial state partons lead to additional DPS contributions 
which have different importance for different nuclei.
Whereas the MPI model of \pythia partially accounts\footnote{Namely, it accounts only for a gluon
  splitting of a type $g\rightarrow q \bar{q}$, for details see
  \cite{Sjostrand:2004pf}.} for processes shown in
Fig~\ref{f:dps_in_pA_1v2}, incorporation of such terms in the Strikman
$\&$ Treleani framework is a non-trivial task, see
\cite{Diehl:2017kgu} and \cite{Diehl:2017wew}.
\begin{table}
\begin{tabular}{ | c | c | c | c | c |}
\hline
  Nucleus                & Angantyr SDTries = 1 & Angantyr SDTries = 2 \\\hline
  \hline
$\prescript{4}{}{{\rm He}}$   & 	1.12	 & 	1.12	 \\\hline
$\prescript{6}{}{{\rm Li}}$   & 	1.18	 & 	1.18	 \\\hline
$\prescript{12}{}{{\rm C}}$   & 	1.34	 & 	1.36 \\\hline
$\prescript{16}{}{{\rm O}}$   & 	1.43	 & 	1.44	 \\\hline
$\prescript{63}{}{{\rm Cu}}$  & 	2.03	 &  	2.03	 \\\hline
$\prescript{129}{}{{\rm Xe}}$ & 	2.46	 &  	2.49	 \\\hline
$\prescript{197}{}{{\rm Au}}$ & 	2.80	 & 	2.80	 \\\hline
$\prescript{208}{}{{\rm Pb}}$ &	2.82	 & 	2.84	 \\\hline
\end{tabular}
\caption{\pythia: predictions for enhancement factor for DPS
  in p$A$ collisions at \mbox{$\sqrt{S_{NN}} = 5$~TeV} ($10^7$ \pythia calls). }
\label{tab:heavy_nuclei_dist_2}
\end{table}

\begin{table}
\begin{tabular}{ | c | c | c | c | c |}
\hline
Nucleus  & Angantyr SDTries = 1  	& Strikman $\&$ Treleani  \\\hline
\hline
$\prescript{4}{}{{\rm He}}$   & 1.12 & 1.21 		 \\\hline
$\prescript{6}{}{{\rm Li}}$   & 1.18	& 1.30		 \\\hline
$\prescript{12}{}{{\rm C}}$   & 1.34	& 1.49		 \\\hline
$\prescript{16}{}{{\rm O}}$   & 1.43	& 1.58		 \\\hline
$\prescript{63}{}{{\rm Cu}}$  & 2.03	& 2.12 		 \\\hline
$\prescript{129}{}{{\rm Xe}}$ & 2.46	& 2.51  		 \\\hline
$\prescript{197}{}{{\rm Au}}$ & 2.80	& 2.78  		 \\\hline
$\prescript{208}{}{{\rm Pb}}$ & 2.82	& 2.82		 \\\hline
\end{tabular}
\caption{\pythia: predictions for enhancement factor
  $\sigma^{\rm DPS}_{\pA} / {A} \sigma^{\rm DPS}_{\rm pp}$ at
  $\sqrt{S_{NN}} = 5$~TeV ($10^7$ \pythia calls). In Strikman $\&$
  Treleani model we set $\sigma_{eff} = 11.3 \, {\rm mb}$.}
\label{tab:angantyr_vs_theory}
\end{table}

\begin{figure}[th]
\centering
\begin{minipage}[h]{1.0\linewidth}
\center{\includegraphics[width=0.6\linewidth]{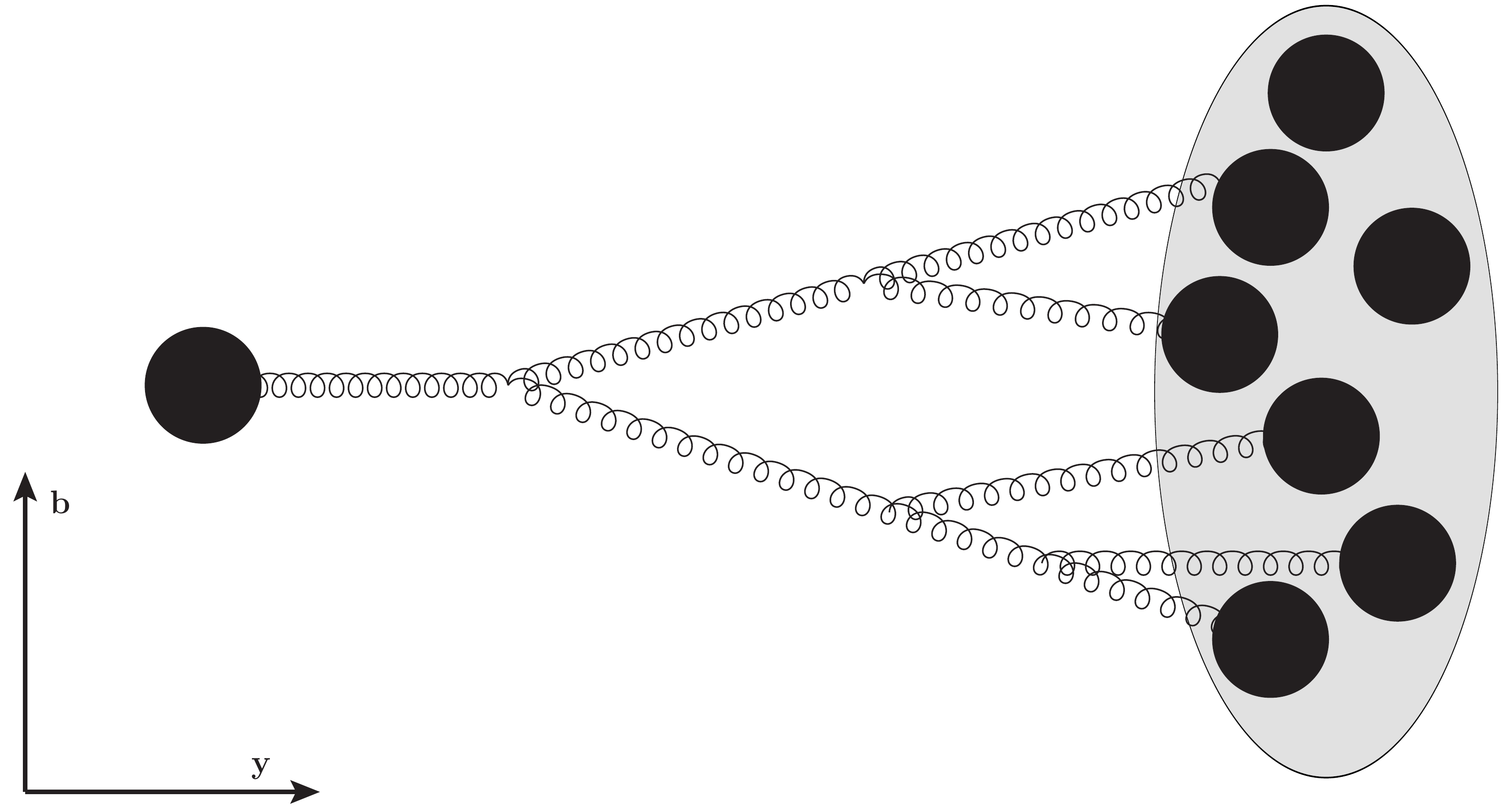}} 
\end{minipage}
\caption{A schematic representation of a BFKL evolution of a gluon
  cascade in rapidity--impact-parameter space.}
\label{f:DIPSY_BFKL_cascade}
\end{figure}

In the previous simulations we were triggering on events with at least
four jets with \mbox{$p_\perp \, \textgreater\, 20$~GeV} without
imposing any cuts on their rapidities. However, it is known that
activity in p$A$ collisions depends on rapidity of produced particles
in a non-trivial way. Namely, as it was observed by the first time by
Busza \textit{et al.} \cite{Busza:1975df}, the charged multiplicity
distribution $dN_{ch} / d\eta$ in p$A$ collisions grows for the
negative values of $\eta$ (assuming that the nucleus $A$ is located in
the negative direction of the $\eta$-axis). There are several
explanation of this phenomenon.  For example, it can be explained by a
BFKL evolution of a gluon cascade in rapidity--impact-parameter space,
as it is sketched in Fig.~\ref{f:DIPSY_BFKL_cascade}, where a
probability to have several absorptive interactions grows in a
direction of a nucleus.  The same result can be explained by the
original non-perturbative ``wounded nucleon model''
\cite{Bialas:1976ed, Nikolaev:1978at, Nikolaev:1981dh} (which is also
the basis of the Fritiof program).  These effects are also
implemented in the the \angantyr model of p$A$ collisions which, to
some extend, can be seen as a perturbative version of the ``wounded
  nucleon model'' which includes MPIs produced according to
Eq.~\eqref{eq:n_sub_col_SD_at_y} with modifications described in
Section \ref{s:angantyr_model}. As it is shown in
\cite{Bierlich:2018xfw} the \angantyr model correctly describes the
aforementioned enhancement.  Since the particle production in
\angantyr relies on the MPI model of the \pythia event generator the
charged multiplicity distribution should be correlated with production
of \mbox{(mini-)jets}. More precisely, the growth of charged
multiplicity $d N_{ch} / d\eta$ for negative $\eta$ values in
\angantyr model is inextricably connected with growth of a number of
sub-scatterings in a given event, see \mbox{Eq.
  \eqref{eq:n_sub_col_SD_at_y}} and
\mbox{Eq.~\eqref{eq:n_sub_col_pp}}. Therefore, it is natural to assume
that in the \angantyr model probability to generate an event of a
\mbox{DPS~II} type will depend on $\eta$ in a way similar to a
$d N_{ch} / d\eta$ distribution. In order to check this we evaluate
$\sigma^{\rm DPS}_{\pA} / {A} \sigma^{\rm DPS}_{\rm pp}$ for events
with at least four jets with \mbox{$p_\perp \textgreater 20$~GeV} and
at least one jet with a pseudo rapidity value smaller\footnote{In our
  simulations we choose a pseudo rapidity axis to run in direction
  from a nucleus to a proton.} than a certain value
$\eta_{cut}$. Obviously, additional $\eta$ cuts will reduce the total
DPS cross section in $\rm pp$ and $\pA$ collisions. Nevertheless, one
could expect that the total DPS cross section in the p$A$ case will
decrease much slower than corresponding one in the pp case. As a
consequence, the enhancement factor
$\sigma^{\rm DPS}_{\pA} / {A} \sigma^{\rm DPS}_{\rm pp}$ will grow
since in the \angantyr model probability to generate a processes of
DPS~II increases at small negative values of $\eta$.

\begin{figure}
\begin{minipage}[h]{1.0\linewidth}
\center{\includegraphics[width=1.0\linewidth]{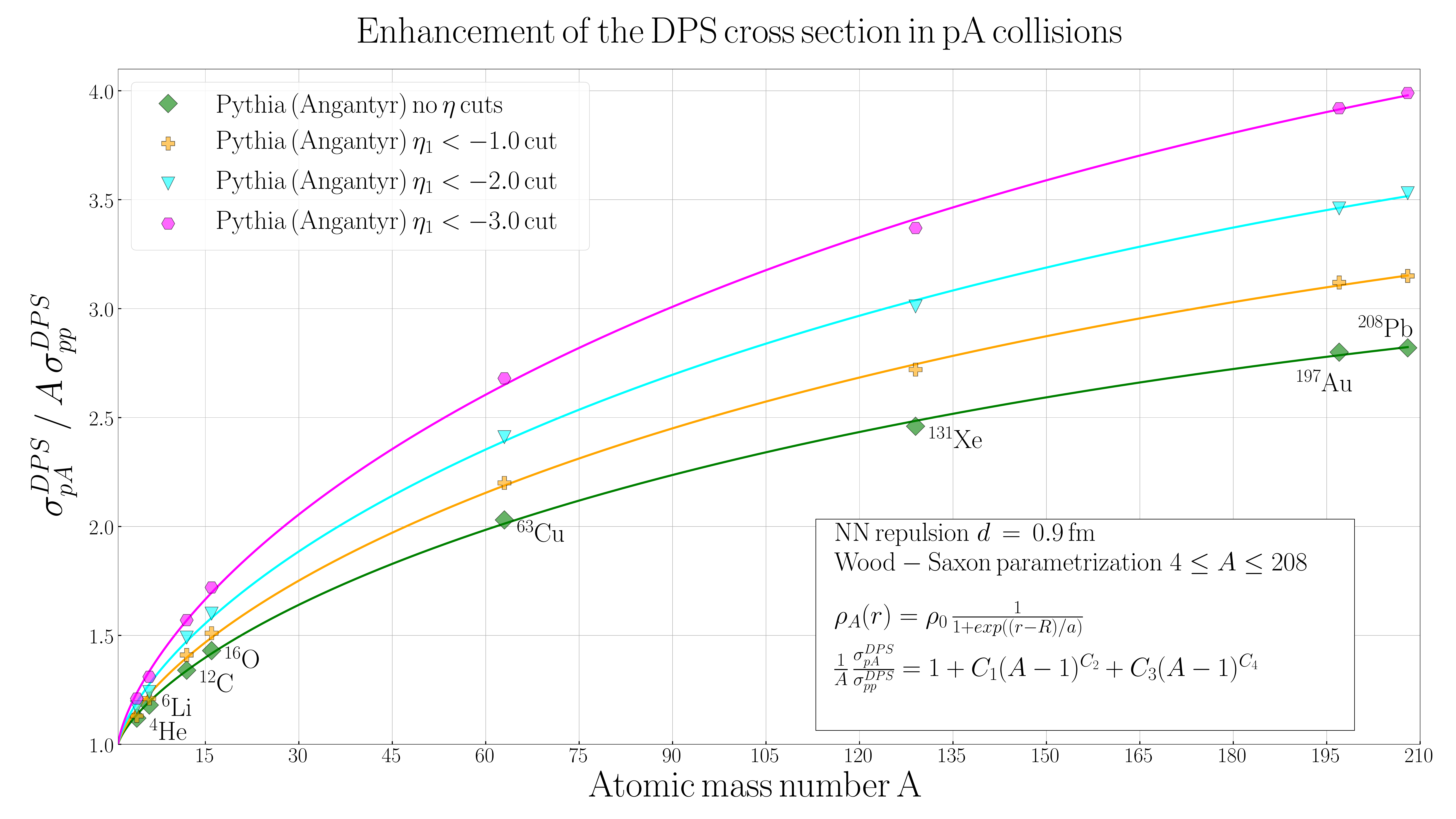}}
\end{minipage}
\cprotect\caption{Dependence of the enhancement factor
  $\sigma^{\rm DPS}_{\pA} / {A} \sigma^{\rm DPS}_{\rm pp}$ on
  $\eta$ cuts. Predictions of \pythia (\angantyr). Here orange, cyan
  and magenta curves correspond to four-jet DPS production with at
  least one jet with $\eta$ smaller than $-1$, $-2$ and $-3$
  correspondingly.}
\label{f:Angantyr_DPS_rapidity_cuts}
\end{figure}

The results are presented in Fig.~\ref{f:Angantyr_DPS_rapidity_cuts}.
In order to study how the DPS enhancement factor
$\sigma^{\rm DPS}_{\pA} / {A} \sigma^{\rm DPS}_{\rm pp}$ depends on
rapidity cuts we have used the same set-up as before but with
additional cuts $\eta_{cut} = -1$, $\eta_{cut} = -2$ and
$\eta_{cut} = -3$. We see that indeed the ratio
$\sigma^{\rm DPS}_{\pA} / {A} \sigma^{\rm DPS}_{\rm pp}$ demonstrates
a strong dependence on the value of $\eta_{cut}$.  The experimental
verification of the growth of the DPS enhancement factor
$\sigma^{\rm DPS}_{\pA} / {A} \sigma^{\rm DPS}_{\rm pp}$ due to the
additional rapidity cut predicted by the \angantyr model could, in
principle, provide a better way to control the fraction of double
absorptive processes shown in Fig.~\ref{f:dps_in_pA_contributions_Angantyr} b).

\section{\label{s:conclusions} Conclusions}

We have demonstrated that the \angantyr model of p$A$ collisions in
\pytppp predicts an $A$-dependence of a DPS enhancement
factor $\sigma^{\rm DPS}_{\pA} / {A} \sigma^{\rm DPS}_{\rm pp}$
which agrees with the one predicted in a pioneering work of Strikman and
Treleani \cite{Strikman:2001gz} at a qualitative level.  This result
can be seen as an additional validation of the \angantyr's approach to
double absorbtive processes described in
Section~\ref{s:angantyr_model}.  From the other side, a correct
$A$-dependence means that, apart from ``standard'' applications,
one can use Angantyr for standalone studies of DPS in p$A$
collisions. In this case a potential user can benefit not only from
evaluation of a total cross section, but also from the most of entire
\pythia machinery like initial and final state radiation, colour
reconnections \textit{etc.} Furthermore the availability of a full
event generator will allow for a realistic estimate of the effects of
the underlying event and other issues associated with the experimental
measurements of jets.

We also have studied how (pseudo) rapidity cuts affect the number of
MPIs in a given event and therefore a behaviour of
$\sigma^{\rm DPS}_{\pA} / {A} \sigma^{\rm DPS}_{\rm pp}$.  The
growth of
$\sigma^{\rm DPS}_{\pA} / {A} \sigma^{\rm DPS}_{\rm pp}$ is a
natural consequence of (pseudo) rapidity dependence of activity in p$A$
collisions built into the \angantyr model. This behaviour was inspired
by a DIPSY model and is essential to get a qualitative agreement with
available experimental data on p$A$ collisions, see
\cite{Bierlich:2018xfw}.

We also argue that, due to the various conceptual differences between
\angantyr and Strikman $\&$ Treleani models one should not expect to
get an exact agreement between their predictions. A complexity of the
problem of DPS in p$A$ collisions requires a detailed study of various
non-trivial effects like partonic correlations, cold nuclear matter
effects and additional DPS contributions, as it was pointed out in
\cite{Blok:2012jr}.  Therefore, in the absence of experimental studies of DPS in p$A$
collisions, a comparison between predictions of \angantyr and improved
Strikman $\&$ Treleani model may help us to identify key ingredients essential for
correct modelling of DPS in p$A$ collisions. Recently, the improved model of  Strikman $\&$ Treleani 
was proposed by Alvioli \textit{et al} \cite{Alvioli:2019kcy}.  In particular it 
accounts for colour fluctuation effects and allows to compute the DPS cross section as a function of centrality. The latter is 
crucial for the experimental studies of the DPS phenomena in p$A$ collisions. Therefore, we argue that, in the
absence of experimental measurements of DPS in p$A$ collisions, the detailed comparison between \angantyr 's predictions
can be beneficial for better understanding of the DPS phenomena in $\pA$ collisions.

\begin{acknowledgments}
The work of OF has received funding from the European Union's Horizon
2020 research and innovation programme as part of the Marie
Sk\l{}odowska-Curie Innovative Training Network MCnetITN3 (grant
agreement no. 722104) and partially by the Deutsche
Forschungsgemeinschaft (DFG) through the Research Training Group ``GRK
2149: Strong and Weak Interactions - from Hadrons to Dark Matter'', 
by the curiosity-driven grant ``Using jets to challenge the Standard Model of particle physics'' 
from Universit\`{a} di Genova, and by the Swedish Research Council, contracts number 2016- 03291,
2016-05996 and 2017-0034 . OF thanks all members of the theoretical
particle physics group of the Lund University for warm hospitality and
friendly atmosphere, and in particular acknowledges Christian
Bierlich, Johannes Bellm, G\"osta Gustafson, Anna Kulesza, Harsh Shah
and Torbj\"orn Sj\"ostrand for useful and fruitful discussions.

All diagrams in this paper were created with the \mbox{JaxoDraw} code
\cite{Binosi:2003yf}.  In order to perform numerical computations
within Strikman $\&$ Treleani framework we have used numerical
integration routines from GSL library \cite{GSL:manual}, in particular
those based upon VEGAS algorithm by Lepage \cite{Lepage:1977sw}.  All
figures in this paper  were
created with the Matplotlib library \cite{Hunter:2007ouj}. For the fitting
purposes SciPy \cite{Virtanen:2019joe} and NumPy \cite{NumPy} libraries were
used. For our computations and simulations involving PDFs we were
using LHAPDF6 library \cite{Buckley:2014ana} and a central value of
MSTW2008 LO PDF set \cite{Martin:2009iq}.
\end{acknowledgments}
\clearpage
\onecolumngrid
\appendix
\section{\label{appendix:pythia_settings} \protect\pythia settings}
\begin{table}[!ht]
  \begin{tabular}{ | l | c |}
    \hline
    \pythia settings						& value \\ \hline
    \hline
    Random:setSeed 						& on	\\ \hline
    HardQCD:all  						& on	\\ \hline
    PartonLevel:mpi 						& on	\\ \hline
    PartonLevel:Remnants 				& on	\\ \hline
    Check:event 							& on	\\ \hline
    PartonLevel:isr 						& off	\\ \hline
    PartonLevel:fsr 						& off	\\ \hline
    ColourReconnection:reconnect 		& off	\\ \hline
    HadronLevel:all 						& off	\\ \hline
    Beams:idA 							& 2212	\\ \hline
   Beams:idB 							& 1000020040 (as an example for $\prescript{4}{}{{\rm He}}$ 	) \\ \hline
    Beams:eA 							& 4000~GeV\\ \hline
    Beams:eB 							& 1570~GeV\\ \hline
    Beams:frameType 						& 2 \\ \hline
   PDF:pSet 							& LHAPDF6:MSTW2008lo68cl \\ \hline
    PhaseSpace:pTHatMin 					& 20.0~GeV \\ \hline
    SigmaProcess:renormScale2 			& 2 \\ \hline
    SigmaProcess:factorScale2 			& 2	 \\ 
    \hline
  \end{tabular}
  \caption{The \pythia settings used for the presented predictions.}
  \label{tab:pythia_flags} 
\end{table}

\twocolumngrid

\bibliography{journal}

\end{document}